\documentclass[preprint,superscriptaddress,showpacs,tightenlines]{revtex4}
\usepackage{graphicx,bm}
\usepackage{amssymb}
\usepackage{amsmath}

\input{tcilatex}

\begin{document}

\title{ Fermi surface contours obtained from STM images around surface point
defects}
\author{N V Khotkevych-Sanina}
\affiliation{B.I. Verkin Institute for Low Temperature Physics and Engineering, National
Academy of Sciences of Ukraine, 47, Lenin Ave., 61103, Kharkov,Ukraine}
\author{Yu A Kolesnichenko}
\affiliation{B.I. Verkin Institute for Low Temperature Physics and Engineering, National
Academy of Sciences of Ukraine, 47, Lenin Ave., 61103, Kharkov,Ukraine}
\author{J M van Ruitenbeek}
\affiliation{Kamerlingh Onnes Laboratorium, Universiteit Leiden, Postbus 9504, 2300
Leiden, The Netherlands}

\begin{abstract}
We present a theoretical analysis of the standing wave patterns in STM
images, which occur around surface point defects. %We address the
%problem of the reconstruction of the Fermi contours for the surface electrons.
We consider arbitrary dispersion relations for the surface states and
calculate the conductance for a system containing a small-size tunnel
contact and a surface impurity. We find rigorous theoretical relations
between the interference patterns in the real-space STM images, their
Fourier transforms and the Fermi contours of two-dimensional electrons. We
propose a new method for reconstructing Fermi contours of surface electron
states, directly from the real-space STM images around isolated surface
defects.
\end{abstract}

\maketitle

\section{Introduction}

Images obtained by scanning tunneling microscope (STM) on flat metal
surfaces commonly display standing waves related to electron scattering by
surface steps and single defects \cite{Crommie} (for a review see \cite%
{Petersen,Simon}). The physical origin of the interference patterns in the
constant-current STM images is the same as that of Friedel oscillations in
the electron local density of states (LDOS) in the vicinity of a scatterer
\cite{Friedel}. It is due to quantum interference between incident electron
waves and waves scattered by the defects. Study of the standing wave pattern
provides information on the defect itself and on the host metal. From the
images the Fermi surface contours (FC) for two-dimensional (2D) surface
states \cite%
{Petersen,Simon,Petersen2,Vonau,Sprunger,Petersen3,Fujita,Briner,Petersen4,Kralj}
and bulk Fermi surfaces can be studied \cite%
{Avotina06,Avotina08,Weismann,Lounis,AKR}.

Let us consider the question of the nature of the contour that we see in a
real space STM\ image. Can it be interpreted directly as the FC or some
contour related with it? For isotropic (circular) FC the answer is obvious -
the period of the conductance oscillations $\Delta r=2\pi /2\kappa _{\mathrm{%
F}}=\mathrm{const}$ is set by twice the 2D Fermi wave vector, $2\kappa _{%
\mathrm{F}}$. In the case of an anisotropic dispersion relation in real
space the electrons move in the direction given by their velocity $\mathbf{v}%
_{\mathbf{\kappa }}$, which needs not be parallel to the wave vector $%
\mathbf{\kappa }.$ We expect that, similar to the problem of subsurface
defects in the bulk \cite{Avotina06,Avotina08}, the period of the real space
oscillatory pattern is $\Delta r=2\pi /2\mathbf{\kappa }_{\mathrm{F}}\mathbf{%
n}_{0},$ were $\mathbf{n}_{0}$ is the 2D unit vector pointing from the
defect to the position of the tip apex, and $\mathbf{\kappa }_{\mathrm{F}}$
is the Fermi wave vector, the magnitude of which depends on its direction.
Thus, in the STM image we observe a curve shaped by the projection of the
wave vector $\mathbf{\kappa }_{\mathrm{F}}$ on the normal to the FC. In the
case of a large anisotropy this contour may be very different from the FC
itself. In Refs.~\cite%
{Petersen,Simon,Petersen2,Vonau,Sprunger,Petersen3,Fujita,Briner,Petersen4,Kralj}
Fourier transforms (FT) of STM images of wave patterns around surface point
defects were interpreted as FC. We are not aware of any rigorous
mathematical justification of this procedure. Is there an other way for
reconstructing the true FC from real space STM images? The answer to this
question is the main aim of this paper.

The STM theory used most frequently by experimentalists is the approach of
Tersoff and Hamann \cite{TH}. Their theoretical analysis of the tunnel
current is based on Bardeen's approximation \cite{Bardeen}, in which a
tunneling matrix element is calculated using the decay of the wave functions
of the two individual (isolated) electrodes inside the barrier. For the STM
tip they adopt a model of angle-independent wave functions and the surface
states are described by Bloch wave functions, which decay exponentially
inside the tunnel barrier. The authors of Ref. \cite{TH} found that the STM
conductance is directly proportional to the electron LDOS $\rho \left(
\mathbf{r}\right) $ at a point $\mathbf{r}=\mathbf{r}_{0}$, at the position
of the contact. In the same sppirit, the influence on the STM conductance of
adatoms or defects embedded into the sample surface is usually described by
their influence on the 2D LDOS. This was used, for example, to explain the
observation of a "quantum mirage" in "quantum corrals" \cite{corral} in
terms of a free-electron approximation, and for the interpretation of the
anisotropic standing Bloch waves observed on Be surfaces \cite{Briner,Simon}%
. In spite of the large number of theoretical works dealing with STM theory
(for review see \cite{Hofer,Blanco}), a number of questions of the
theoretical description of anisotropic standing wave patterns in STM images
remains poorly described.

The main new points of present paper are: 1) In an approximation of free
electrons with an arbitrary anisotropic dispersion law the quantum electron
tunneling through a small contact into Shockley-like two-dimensional surface
states is considered theoretically. In the framework of a model of an
inhomogeneous $\delta$-like tunnel barrier, Ref.~\cite{Kulik,Avotina05}, we
obtain analytical formulas for the conductance $G$ of the contact in the
presence of a single defect incorporated in the sample surface. 2) We
formulate a rigorous mathematical procedure for the FC reconstruction from
real space images of conductance oscillations around surface point-like
defects in the terms of a support function of a plane curve (see, for
example, \cite{Differential Geometry}) .

The organization of this paper is as follows. The model that we use to
describe the contact, and the basic equations are presented in Sec.~II. In
Sec.~III the differential conductance is found on the basis of a calculation
of the probability current density through the contact. Sec.~IV presents the
mathematical procedure of reconstruction of FC in the momentum space from
the real space image. In Sec.~V we conclude by discussing the possibilities
for exploiting these theoretical results for interpretation of STM
experiments. In App. I the method for obtaining a solution of the Schr\"{o}%
dinger equation is described, and in App. II we find the asymptote of the 2D
electron Green's function for large values of the coordinates, which is
necessary to describe the conductance oscillations at large distances
between the tip and the defect.

\section{Model of the system and basic equations.}

The STM tip and a conducting surface form an atomic size tunnel contact. The
STM image is obtained from the height profile while maintaining the tunnel
current $I$ constant, or from the differential conductance $G=dI/dV)$
measured as a function of the lateral coordinates. Such dependencies are a
kind of electronic "maps" of the surface, thereby they show a variety of
defects situated on the metal's surface (adsorbed and embedded impurities,
steps, etc.) We focus our attention on studying the shape of contours of
oscillatory patterns around a single point defect. These concentric contours
with the center on the defect (see Fig.1) are minima and maxima of the
oscillatory dependence of the conductance on the lateral coordinates.
\begin{figure}[tbp]
\includegraphics[width=10cm,angle=0]{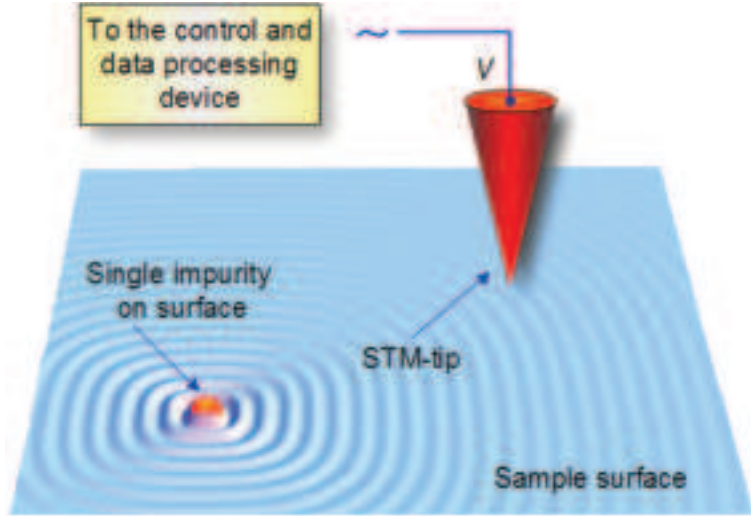}
\caption{Schematic setup for measurements with a STM. A standing wave
pattern arising around a single impurity on the surface is shown.}
\label{Fig1}
\end{figure}
\textbf{\ }

In Ref.~\cite{Avotina05} it was proposed to model the STM experiments by an
inhomogeneous infinitely thin tunnel barrier. The important simplification
offered by this model is the replacement of the three-dimensional
inhomogeneous tunnel barrier in a real experimental configuration by a
two-dimensional one. In the present paper we use this model to describe the
interference pattern around a point defect resulting from electron surface
states having an anisotropic FC. Instead of a description by means of simple
Bloch waves we consider quasiparticles \cite{LAK} for conduction electrons
having arbitrary dispersion relations.

The model that we consider is presented in Fig.~2. Two conducting
half-spaces are separated by an infinitely thin insulating interface at $%
z=0, $ the potential barrier $U\left( \mathbf{r}\right) $ in the plane of
which we describe by Dirac delta function $U\left( \mathbf{r}\right)
=U_{0}f\left( \mathbf{\rho }\right) \delta \left( z\right) .$ A
dimensionless function $f\left( \mathbf{\rho }\right) $ describes a barrier
inhomogeneity in the plane $\mathbf{\rho }=\left( x,y\right) .$ This
inhomogeneity simulates the STM tip and provides the path for electron
tunneling through a bounded region of scale $a\lesssim \lambda _{\mathrm{F}}$
($a$ is the characteristic radius of the contact, $\lambda _{\mathrm{F}}$ is
the electron Fermi wave length), i.e. the function $f\left( \mathbf{\rho }%
\right) $ must have the property
\begin{equation}
f\left( \mathbf{\rho }\right) =\left\{
\begin{array}{c}
\sim 1,\text{ }\rho \lesssim a \\
\rightarrow \infty ,\text{ }\rho \gg a%
\end{array}%
\right. .  \label{f}
\end{equation}%
Simple examples of such function are $f\left( \mathbf{\rho }\right) =\exp
\left( \rho ^{2}/a^{2}\right) $ and $1/f\left( \mathbf{\rho }\right) =\Theta
\left( a-\rho \right) ,$ were $\Theta \left( x\right) $ is the Heaviside
step function. The latter function corresponds to a model with a circular
orifice of radius $a$ in an otherwise impenetrable interface.

Shockley-like surface states are included in the model by means of a surface
potential $V_{\mathrm{sur}}\left( z\right) $ in the half-space $z>0$. The
potential $V_{\mathrm{sur}}\left( z\right) $ and the barrier at $z=0$ form a
quantum well which localizes electrons near the surface. A specific form of
the function $V_{\mathrm{sur}}\left( z\right) $ is not important for us. It
is enough to assume that $V_{\mathrm{sur}}\left( z\right) $ is analytic
monotonous function such that it permits the existence of one and only one
surface state in the region $z>0$ below the Fermi energy $\varepsilon _{%
\mathrm{F}}$. The surface state localization length $l$ is assumed to be
much larger than the characteristic contact diameter, $a\ll l.$

A a single point-like defect is placed in a point $\mathbf{r}_{0}=\left(
\mathbf{\rho }_{0},z_{0}>0\right) $ in the vicinity of the interface at $%
z=0. $ The electron scattering by the defect we describe by a short range
potential $D\left( \mathbf{r}\right) =gD_{0}\left( \mathbf{r-r}_{0}\right) $
localized within a region of characteristic radius $r_{D}$, and in the
half-space $z>0$, around the point $\mathbf{r}_{0}=\left( \mathbf{\rho }%
_{0},z_{0}\right) $, where $g$ is the constant measuring the strength of the
electron interaction with the defect. It satisfies the normalization
condition
\begin{equation}
\int\limits_{-\infty }^{\infty }d\mathbf{r}D_{0}\left( \mathbf{r-r}%
_{0}\right) =1.  \label{D_0_norm}
\end{equation}%
\begin{figure}[tbp]
\includegraphics[width=10cm,angle=0]{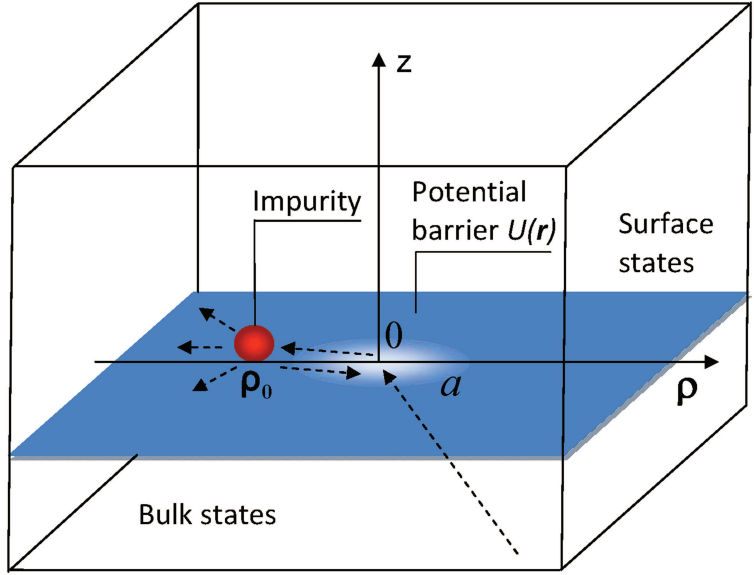}
\caption{Illustration of the model used for the description of STM tunneling
into the surface states. The blue colored region of the interface at $z=0$
separates the tip (lower half) from the sample (upper half). In the center
of the interface at the point $\mathbf{r}=0$ a region is shown having
maximal probability of electron tunneling, which models a contact of
characteristic radius $a$. In the point $\mathbf{r}=\left( \mathbf{\protect%
\rho }_{0},z_{0}\right) $ a point-like defect is situated. Arrows
schematically show semiclassical trajectories of electrons, for the bulk
states in the half- space $z<0$, and for the surface electrons at $z>0.$ }
\label{fig1}
\end{figure}

We do not specify the concrete form of the potential $D\left( \mathbf{r}%
\right).$ The specific form affects the amplitude and phase of the
conductance oscillations but it does not change their period, which is the
main subject of interest for us. We assume the following general properties
for this function: 1) The potential is repulsive and electron bound states
near the defect are absent. 2) The constant of interaction $g$ in the
potential $D\left( \mathbf{r}\right) $ is small such that Born's
approximation for waves scattered by the defect is applicable. 3) The
effective radius $r_{D}$ of the potential $D\left( \mathbf{r}\right) $ is
small enough $\kappa _{\mathrm{F}}r_{D}\ll 1$ for the scattering to be
described in the s-wave approximation. All of the listed conditions can be
easily satisfied in experiments.

In order to obtain an analytical solution of the Schr\"{o}dinger equation
and calculate the electric current in what follows we use a simplified model
for the anisotropic dispersion law $\varepsilon \left( \mathbf{k}\right) $
for the charge carriers%
\begin{equation}
\varepsilon \left( \mathbf{k}\right) =\varepsilon _{2D}\left( \mathbf{\kappa
}\right) +\frac{\hbar ^{2}k_{z}^{2}}{2m_{z}},  \label{eps(k)}
\end{equation}%
where $\mathbf{\kappa }$ is the 2D electron wave vector in the plane of
interface, $\varepsilon _{2D}\left( \mathbf{\kappa }\right) \ $is an
arbitrary function describing the energy spectrum of the surface states, and
$\ m_{z}$ is the effective mass characterizing the electron motion along the
normal to the interface.

The wave function $\psi $ satisfies the Schr\"{o}dinger equation
\begin{equation}
\left( \varepsilon _{2D}\left( \frac{\hbar \partial }{i\partial \mathbf{\rho
}}\right) +\frac{\hslash ^{2}\partial ^{2}}{2m_{z}\partial z^{2}}\right)
\psi \left( \mathbf{r}\right) +\left[ \varepsilon -U\left( \mathbf{r}\right)
-D\left( \mathbf{r}\right) -V_{\mathrm{sur}}\left( z\right) \Theta \left(
z\right) \right] \psi \left( \mathbf{r}\right) =0,  \label{Schrod}
\end{equation}%
with $\varepsilon $ the electron energy. At the interface $z=0$ the function
$\psi \left( \mathbf{r}\right) $ satisfies the boundary conditions for
continuity of the wave function
\begin{equation}
\psi \left( \mathbf{\rho },+0\right) =\psi \left( \mathbf{\rho },-0\right) ,
\label{equal}
\end{equation}%
and jump of its derivative
\begin{equation}
\psi _{z}^{\prime }\left( \mathbf{\rho },+0\right) -\psi _{z}^{^{\prime
}}\left( \mathbf{\rho },-0\right) =\frac{2m_{z}U_{0}}{\hbar ^{2}}f\left(
\mathbf{\rho }\right) \psi \left( \mathbf{\rho },0\right) ,  \label{jump_wf}
\end{equation}%
and the condition for the decay of the surface state wave function in the
classically forbidden region%
\begin{equation}
\psi \left( \mathbf{\rho },z\rightarrow \infty \right) \rightarrow 0.
\label{inf}
\end{equation}%
For $z\rightarrow -\infty $ the solution $\psi \left( \mathbf{r}\right) $ of
Eq. (\ref{Schrod}) must describe waves emanating from the contact \cite%
{Avotina06}%
\begin{equation}
\psi \left( \left\vert \mathbf{r}\right\vert \rightarrow \infty ,z<0\right)
\sim \frac{\exp \left( i\mathbf{kn}r\right) }{r},  \label{-inf}
\end{equation}%
where $\mathbf{n}$ is a unit vector directed along the velocity vector $%
\mathbf{v}_{\mathbf{k}}=\partial \varepsilon \left( \mathbf{k}\right)
/\partial \mathbf{k}.$

In order to calculate the tunnel current at small applied voltage $V$ ($%
eV\ll \varepsilon _{\mathrm{F}})$ we must find the wave function $\psi _{%
\mathrm{tr}}\left( \mathbf{\rho },z\right) $ for electrons transmitted
through the tunnel barrier. By means of this function the density of current
flow and the total current in the system can be calculated. At zero
temperature it is enough to consider one direction of tunneling. For
definiteness we select the sign of the voltage such that the tunneling
occurs from the surface states at $z>0$ into the bulk states at $z<0$ (see
Fig.3). The total current $I$ can be found by the integration over the wave
vectors ${\mathbf{\kappa }}$ of the surface states and integration over
coordinate $\mathbf{\rho }$ in the plane $z=\mathrm{const.}\neq 0$ in the
half - space $z<0$
\begin{equation}
I{=-\frac{e^{2}\hbar L_{x}L_{y}V}{2\pi ^{2}m_{z}}\int\limits_{-\infty
}^{\infty }d\mathbf{\kappa }\int\limits_{-\infty }^{\infty }d\mathbf{\rho }%
\func{Im}\left[ \psi _{\mathrm{tr}}^{\ast }\left( \mathbf{\rho },z\right)
\frac{\partial }{\partial z}\psi _{\mathrm{tr}}\left( \mathbf{\rho }%
,z\right) \right] }\frac{\partial {f_{\mathrm{F}}\left( \varepsilon \right) }%
}{\partial \varepsilon }.  \label{I}
\end{equation}%
Here $L_{x,y}$ is the size of the sample in corresponding direction.
\begin{figure}[tbp]
\includegraphics[width=10cm,angle=0]{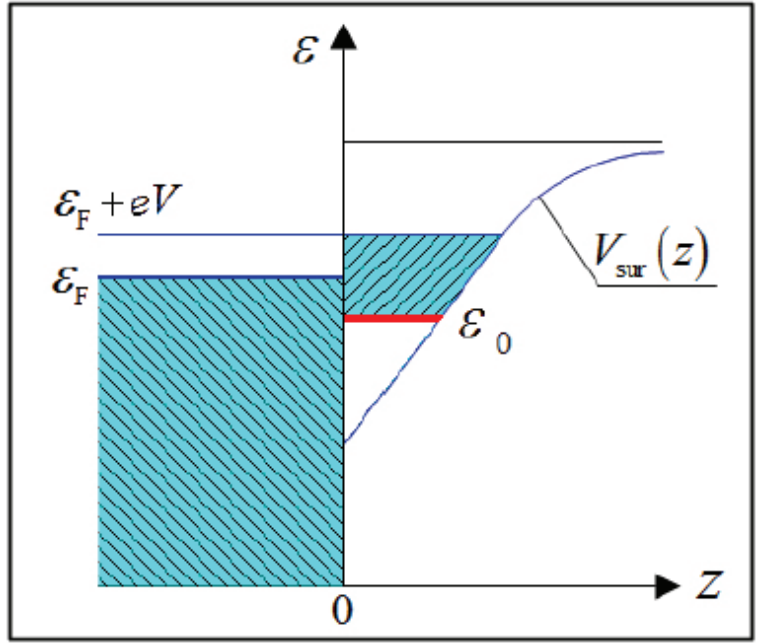}
\caption{Illustration of the occupied energy bands near the interface. The
applied bias $eV$ makes it possible for the electrons to tunnel from surface
states into bulk states of the STM tip. }
\end{figure}

The procedure for the solution of the Schr\"{o}dinger equation (\ref{Schrod}%
) with boundary conditions (\ref{equal})-(\ref{-inf}) is presented in
Appendix I, were the wave function $\psi _{\mathrm{tr}}\left( \mathbf{\rho }%
,z\right) $ (\ref{psi_tr}) is found.

\section{Standing wave pattern in the conductance of a small contact}

Obviously, in the case of small applied bias which we consider in this work,
$\left\vert eV\right\vert \ll \varepsilon _{\mathrm{F}}$, with $\varepsilon
_{\mathrm{F}}$ the Fermi energy, the conductance $G=I/V$ does not depend on
the direction of the current. Substituting the wave function $\psi _{\mathrm{%
tr}}\left( \mathbf{\rho },z\right) $ (\ref{psi_tr}) in Eq. (\ref{I}) after
some integrations we find%
\begin{gather}
G=\frac{e^{2}\hbar ^{5}\left\vert \chi _{0z}^{\prime }\left( +0\right)
\right\vert ^{2}}{32\pi ^{4}m_{z}^{3}U_{0}^{2}}\int\limits_{-\infty
}^{\infty }\int\limits_{-\infty }^{\infty }\frac{d\mathbf{\rho }^{\prime }}{%
f\left( \mathbf{\rho }^{\prime }\right) }\frac{d\mathbf{\rho }^{\prime
\prime }}{f\left( \mathbf{\rho }^{\prime \prime }\right) }%
\int\limits_{0}^{\infty }{d\mathbf{\kappa }}^{\prime }k_{z}^{\prime }\Theta
\left( \varepsilon _{\mathrm{F}}-\varepsilon _{2D}\left( \mathbf{\kappa }%
^{\prime }\right) \right) \cos \left[ {\mathbf{\kappa }}^{\prime }\left(
\mathbf{\rho }^{\prime }\mathbf{-\rho }^{\prime \prime }\right) \right]
\int\limits_{0}^{\infty }{d\mathbf{\kappa }}\cdot  \label{Ggen} \\
\delta \left( \varepsilon _{\mathrm{F}}-\varepsilon _{2D}\left( \mathbf{%
\kappa }\right) -\varepsilon _{0}\right) \left[ \cos \left[ {\mathbf{\kappa }%
}\left( \mathbf{\rho }^{\prime }\mathbf{-\rho }^{\prime \prime }\right) %
\right] +2g\int\limits_{-\infty }^{\infty }d\mathbf{\rho }^{\prime \prime
\prime }\int\limits_{-\infty }^{\infty }dz^{\prime \prime \prime
}D_{0}\left( \mathbf{\rho }^{\prime \prime \prime }\mathbf{-\rho }%
_{0},z^{\prime \prime \prime }-z_{0}\right) \left\vert \chi _{0}\left(
z^{\prime \prime \prime }\right) \right\vert ^{2}\right. \cdot  \notag \\
\left. \cos \left[ {\mathbf{\kappa }}\left( \mathbf{\rho }^{\prime }\mathbf{%
-\rho }^{\prime \prime \prime }\right) \right] \func{Re}G_{2D}^{+}\left(
\mathbf{\rho }^{\prime }\mathbf{-\rho }^{\prime \prime \prime };\varepsilon
_{\mathrm{F}}-\varepsilon _{0}\right) \right] .  \notag
\end{gather}%
Further calculations require explicit expressions for the functions $f\left(
\mathbf{\rho }\right) $ and $D_{0}\left( \mathbf{\rho -\rho }%
_{0},z-z_{0}\right) .$ The integral formula (\ref{Ggen}) can be simplified
for contacts of small radius $a$ and in the limit of a short range $r_{D}$
of the scattering potential. If $\kappa _{F}a\ll 1$ and \ $\kappa
_{F}r_{D}\ll 1,$ where $\kappa _{F}=\frac{1}{\hbar }\sqrt{2m_{z}\left(
\varepsilon _{\mathrm{F}}-\varepsilon _{0}\right) }$ , all functions in (\ref%
{Ggen}) under the integrals, except $f\left( \mathbf{\rho }\right) $ and $%
D_{0}\left( \mathbf{\rho -\rho }_{0},z-z_{0}\right) $, can be taken in the
points $\mathbf{\rho }^{\prime }=\mathbf{\rho }^{\prime \prime }=0,$ $%
\mathbf{\rho }^{\prime \prime \prime }\mathbf{=\rho }_{0}$, which simplifies
(\ref{Ggen}) to,
\begin{equation}
G{=}G_{0}\left[ 1+\frac{2\widetilde{g}}{\left( 2\pi \hbar \right) ^{2}\rho
_{2D}\left( \varepsilon _{\mathrm{F}}-\varepsilon _{0}\right) }\func{Re}%
G_{2D}^{+}\left( \mathbf{\rho }_{0};\varepsilon _{\mathrm{F}}-\varepsilon
_{0}\right) \oint\limits_{\varepsilon _{\mathrm{F}}-\varepsilon
_{0}=\varepsilon _{2D}\left( {\mathbf{\kappa }}\right) }\frac{dl_{{\mathbf{%
\kappa }}}}{v_{{\mathbf{\kappa }}}}\cos {\mathbf{\kappa }}\mathbf{\rho }_{0}%
\right] .  \label{G_a_small}
\end{equation}%
Here%
\begin{equation}
G_{0}=\frac{e^{2}\hbar ^{5}\left\vert \chi _{0z}^{\prime }\left( +0\right)
\right\vert ^{2}}{8m_{z}^{3}U_{0}^{2}}S_{eff}^{2}\rho _{2D}\left(
\varepsilon _{\mathrm{F}}-\varepsilon _{0}\right) \Omega \left( \varepsilon
_{\mathrm{F}}\right)  \label{G_0S}
\end{equation}%
is the conductance of a tunnel point contact between the surface states,
unperturbed by defects, and the bulk states of the tip. Further,
\begin{equation}
\rho _{2D}\left( \varepsilon \right) =\frac{2}{\left( 2\pi \hbar \right) ^{2}%
}\oint\limits_{\varepsilon =\varepsilon _{2D}\left( {\mathbf{\kappa }}%
\right) }\frac{dl_{{\mathbf{\kappa }}}}{v_{{\mathbf{\kappa }}}}
\end{equation}%
is the two-dimensional density of states, where the integration is carried
out over the arc length $l_{{\mathbf{\kappa }}}$ of the constant-energy
contour , $v_{{\mathbf{\kappa }}}=\left\vert \partial \varepsilon
_{2D}\left( {\mathbf{\kappa }}\right) /\hbar \partial {\mathbf{\kappa }}%
\right\vert $ is absolute value of the 2D velocity vector,%
\begin{equation}
\Omega \left( \varepsilon \right) =\frac{\sqrt{2m_{z}}}{\hbar }%
\int\limits_{0}^{\varepsilon }d\varepsilon ^{\prime }\sqrt{\varepsilon
-\varepsilon ^{\prime }}\rho _{2D}\left( \varepsilon ^{\prime }\right) ,
\end{equation}%
\begin{equation}
S_{eff}=\int\limits_{-\infty }^{\infty }\frac{d\mathbf{\rho }}{f\left(
\mathbf{\rho }\right) },  \label{S_eff}
\end{equation}%
is the effective area of the contact, and
\begin{equation}
\widetilde{g}=g\int\limits_{-\infty }^{\infty }d\mathbf{\rho }%
\int\limits_{0}^{\infty }dzD_{0}\left( \mathbf{\rho },z-z_{0}\right)
\left\vert \chi _{0}\left( z\right) \right\vert ^{2}  \label{g}
\end{equation}%
is the effective constant of interaction with the defect for the electrons
belonging to the surface states.

For large distances between the contact and the defect, $\kappa _{\mathrm{F}%
}\rho _{0}\gg 1$, Eq.~(\ref{G_a_small}) can be reduced by using an
asymptotic expression for the Green function, see (\ref{G2D_as}) in Appendix
II. The asymptotic form for $\kappa _{\mathrm{F}}\rho _{0}\gg 1$ of the
integral over $l_{{\mathbf{\kappa }}}$ in Eq.~(\ref{G_a_small}) is the real
part of Eq.~(\ref{G2D_asym}). Under the assumptions listed above the formula
for the oscillatory part of the conductance takes the form
\begin{equation}
\frac{G_{osc}\left( \mathbf{\rho }_{0}\right) }{G_{0}}=\widetilde{g}\left.
\frac{\text{sgn}K(\mathbf{\kappa })\cos (2\mathbf{\kappa \rho }_{0})}{2\rho
_{2D}\left( \varepsilon \right) \hbar ^{2}v_{\mathbf{\kappa }}^{2}\left\vert
K\left( \mathbf{\kappa }\right) \right\vert \rho _{0}}\right\vert _{\mathbf{%
\kappa }=\mathbf{\kappa }\left( \varepsilon _{\mathrm{F}}-\varepsilon
_{0},\phi _{0}\right) },  \label{G_asympt}
\end{equation}%
where $\phi _{0}$ satisfies the stationary phase condition (\ref{stphpoint_2}%
) for $\rho =\rho _{0}.$ We emphasize that the result (\ref{G_asympt}) is
valid, if $\kappa _{\mathrm{F}}a\ll 1$\ $\kappa _{\mathrm{F}}r_{D}\ll 1,$\
and \ $\kappa _{\mathrm{F}}\rho _{0}\gg 1.$\ For example, in actual STM
experiments for surface states of Cu$\left( 111\right) $\ \cite{Crommie} the
Fermi wave vector is $\kappa _{\mathrm{F}}\simeq 0.2\mathring{A}^{-1},$\
while $a$\ and $r_{D}$ are of atomic size $a\simeq r_{D}\simeq 1\mathring{A}$%
. The period of real-space conductance oscillations is $\Delta \rho
_{0}\simeq \pi /\kappa _{\mathrm{F}}\simeq 15\mathring{A}$\ and the distance
over which these oscillations are observable reaches $\rho _{0}\simeq 100%
\mathring{A}.$\ Note that the asymptotic form (\ref{G_asympt}) can be used
to describe experimental data with satisfactory accuracy, with the less
strict requirements of $a$\ and $r_{D}$\ smaller than the Fermi wave length $%
\lambda _{\mathrm{F}}=2\pi /\kappa _{\mathrm{F}}$, and $\kappa _{\mathrm{F}%
}\rho _{0}>1.$

In Ref.~\cite{Khotkevych} the expression for the conductance (\ref{Ggen})
has been found for the special case of an elliptic Fermi surface for the
surface charge carriers. Within that model the conductance oscillations $%
G_{osc}$\ can be evaluated correctly in a wider interval of values for $%
\kappa _{\mathrm{F}}\rho _{0}$, including $\kappa _{\mathrm{F}}\rho
_{0}\lesssim 1$\ (but $\rho _{0}\gg a,r_{D}$). A comparison that result with
the asymptotic formula (\ref{G_asympt}) \ for an elliptic Fermi contour
shows that the relative error in the period of oscillations $\Delta \rho
_{0} $\ determined, as an example, as the distance between the third and
fourth maxima in the dependence $G_{osc}\left( \mathbf{\rho }_{0}\right) $\ (%
\ref{G_asympt}) is about a few percent.

\section{Reconstruction of the Fermi contour from real space STM images.}

Above we have shown that for large distances between the contact and the
defect the period of the oscillations in the conductance is defined by the
function $\mathbf{\kappa }\left( \varepsilon _{\mathrm{F}},\phi _{0}\right)
\mathbf{\rho }_{0}\mathbf{.}$ Taking into account that according Eq.~(\ref%
{stphpoint_2}) in the stationary phase point $\mathbf{\kappa }=\mathbf{%
\kappa }\left( \varepsilon ,\phi _{0}\right) $ the vector $\mathbf{\rho }%
_{0} $ is parallel to the electron velocity, $\mathbf{\rho }_{0}\parallel
\mathbf{v}_{\mathbf{\kappa }}\left( \varepsilon _{\mathrm{F}},\phi
_{0}\right) \mathbf{,}$ i.e.
\begin{equation}
\mathbf{\kappa }\left( \varepsilon _{\mathrm{F}},\phi _{0}\right) \mathbf{%
\rho }_{0}\mathbf{=\kappa n}_{\mathbf{\kappa }}\rho _{0}=h\left( \varepsilon
_{\mathrm{F}},\phi _{0}\right) \rho _{0},
\end{equation}%
were $\mathbf{v}_{\mathbf{\kappa }}/v_{\mathbf{\kappa }}=\mathbf{n}_{\mathbf{%
\kappa }}$ is the unit vector normal to the contour of constant energy $%
\varepsilon _{2D}\left( \mathbf{q}\right) =\varepsilon _{\mathrm{F}%
}-\varepsilon _{0}$ in the point defined by wave vector $\mathbf{\kappa }%
\mathbf{.}$

\begin{figure}[tbp]
\includegraphics[width=10cm,angle=0]{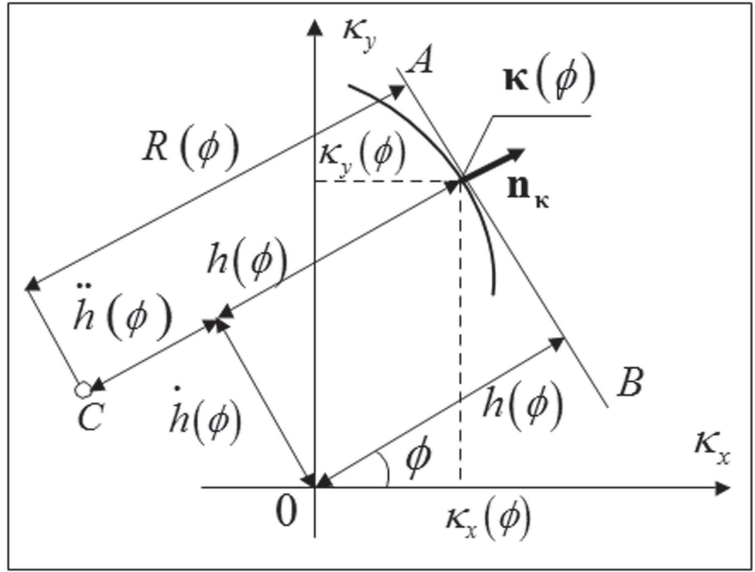}
\caption{Geometric relations between the coordinates $\protect\kappa %
_{x}\left( \protect\phi \right) $ and $\protect\kappa _{y}\left( \protect%
\phi \right) $ of the parametrically defined convex curve and the support
function $h\left( \protect\phi \right) $ and its derivatives $\dot{h}\left(
\protect\phi \right) ,$ $\ddot{h}\left( \protect\phi \right) $ with respect
to the angle $\protect\phi ,$ the radius of curvature $R\left( \protect\phi %
\right) $, and the normal vector $\mathbf{n}_{\protect\kappa }$ at the point
$\mathbf{\protect\kappa }\left( \protect\phi \right) .$ $AB$ is the tangent
to the curve in the point $\mathbf{\protect\kappa }\left( \protect\phi %
\right) $.}
\label{figy}
\end{figure}

By definition $h\left( \phi \right) =\mathbf{\kappa n}_{\mathbf{\kappa }}>0,$
the distance of the tangent from the origin, is the support function for a
convex plane curve \cite{Differential Geometry}. In Fig.~\ref{figy} we
illustrate the geometrical relation between the curve and its support
function. For known $h\left( \phi \right) $ and its first and second
derivatives, $\dot{h}\left( \phi \right) $ and $\ddot{h}\left( \phi \right) $%
, the convex plane curve is given by the parametric equations \cite%
{Differential Geometry},
\begin{eqnarray}
\kappa _{x}\left( \phi \right) &=&h\left( \phi \right) \cos \phi -\dot{h}%
\left( \phi \right) \sin \phi ,  \label{kx} \\
\kappa _{y}\left( \phi \right) &=&h\left( \phi \right) \sin \phi +\dot{h}%
\left( \phi \right) \cos \phi ,  \label{ky}
\end{eqnarray}%
and the radius of curvature $R\left( \phi \right) $ is%
\begin{equation}
R\left( \phi \right) =h\left( \phi \right) +\ddot{h}\left( \phi \right)
\end{equation}%
The curvature is $K\left( \phi \right) =1/R\left( \phi \right) .$ Obviously,
for a circle $\left\vert \mathbf{\kappa }\right\vert $ is constant and the
support function coincides with the circle radius, $h=\kappa .$

Maxima and minima in the oscillatory dependence of the conductance are the
curves of constant phase of the oscillatory functions in Eq.(\ref{G_asympt}%
), $2h(\phi _{0})\rho _{0}=\mathrm{const.}$, and visible contours are
defined by the function%
\begin{equation}
\rho _{0}\left( \phi _{0}\right) =\frac{\mathrm{const.}}{2h\left( \phi
_{0}\right) }.  \label{ro0(phi0)}
\end{equation}

Thus, the Eqs. (\ref{ro0(phi0)}) and (\ref{kx}), (\ref{ky}) in principle
offer the possibility of FC reconstruction from the real space images. For a
non-convex contour $\rho _{0}\left( \phi _{0}\right) $ (\ref{ro0(phi0)}) may
be separated on parts having a constant sign of curvature and for each of
them the above described procedure of reconstruction can be applied.

In order to answer the question what contour is obtained from the Fourier
transform $F(\mathbf{q})$ of an STM image we analyze Eq.~(\ref{G_asympt}).
Although (\ref{G_asympt}) is strictly valid only for $\kappa _{F}\rho
_{0}\gg 1$ in the region $\kappa _{F}\rho _{0}>1$ the difference of the true
period $\Delta \rho _{0}$ of the oscillations from the value $\Delta \rho
_{0}=\pi /h\left( \phi _{0}\right) $ is small, as mentioned for an elliptic
Fermi contour above. Performing the Fourier transform we find%
\begin{gather}
(2\pi )^{2}F\left( \mathbf{q}\right) =\int\limits_{0}^{\infty }d\rho \rho
\int\limits_{0}^{2\pi }d\phi \frac{\cos \left( 2h(\phi )\rho \right) }{\rho }%
\exp \left( iQ\left( \phi \right) \rho \right) =  \label{F} \\
\frac{\pi }{2}\left[ \frac{1}{2\dot{h}\left( \phi _{1}\right) -\dot{Q}\left(
\phi _{1}\right) }+\frac{1}{2\dot{h}\left( \phi _{2}\right) +\dot{Q}\left(
\phi _{2}\right) }\right] -i\int\limits_{0}^{2\pi }d\phi \frac{Q\left( \phi
\right) }{Q^{2}\left( \phi \right) -4h^{2}\left( \phi \right) },  \notag
\end{gather}%
where%
\begin{equation}
Q\left( \phi \right) =\left( q_{x}\cos \phi +q_{y}\sin \phi \right) ,
\label{Q}
\end{equation}%
and $\phi _{1,2}=\phi _{1,2}\left( q_{x},q_{y}\right) $ are the solutions of
the equations%
\begin{equation}
2h\left( \phi _{1,2}\right) =\pm Q\left( \phi _{1,2}\right) .  \label{2h=Q1}
\end{equation}%
The function $F\left( \mathbf{q}\right) $ (\ref{F}) has a singularity when%
\begin{equation}
2\dot{h}\left( \phi _{1,2}\right) =\pm \dot{Q}\left( \phi _{1,2}\right) .
\label{2h=Q2}
\end{equation}%
From Eqs.~(\ref{kx}) and (\ref{ky}) it follows that for $\kappa _{x}$ and $%
\kappa _{y}$ belong to a contour of constant energy,
\begin{gather}
\kappa _{x}\cos \phi +\kappa _{y}\sin \phi =h\left( \phi \right) >0,
\label{h} \\
\kappa _{x}\sin \phi -\kappa _{y}\cos \phi =-\dot{h}\left( \phi \right).
\label{-h'}
\end{gather}
It is easy to see that simultaneous fulfillment of the conditions (\ref%
{2h=Q1}) and (\ref{2h=Q2}) at $\phi =\phi _{1}$ is equivalent to the Eqs.~(%
\ref{h}) and (\ref{-h'}), which are the parametric equations of the constant
energy contour, i.e. the Fourier transform gives the doubled FC of the
surface state electrons. The solution $\phi =\phi _{2}$ of the second
equation corresponds to the reflection symmetry point $-\mathbf{q=}\left(
-q_{x},-q_{y}\right) $ of the 2D Fermi surface.
\begin{figure}[tbp]
\includegraphics[width=15cm,angle=0]{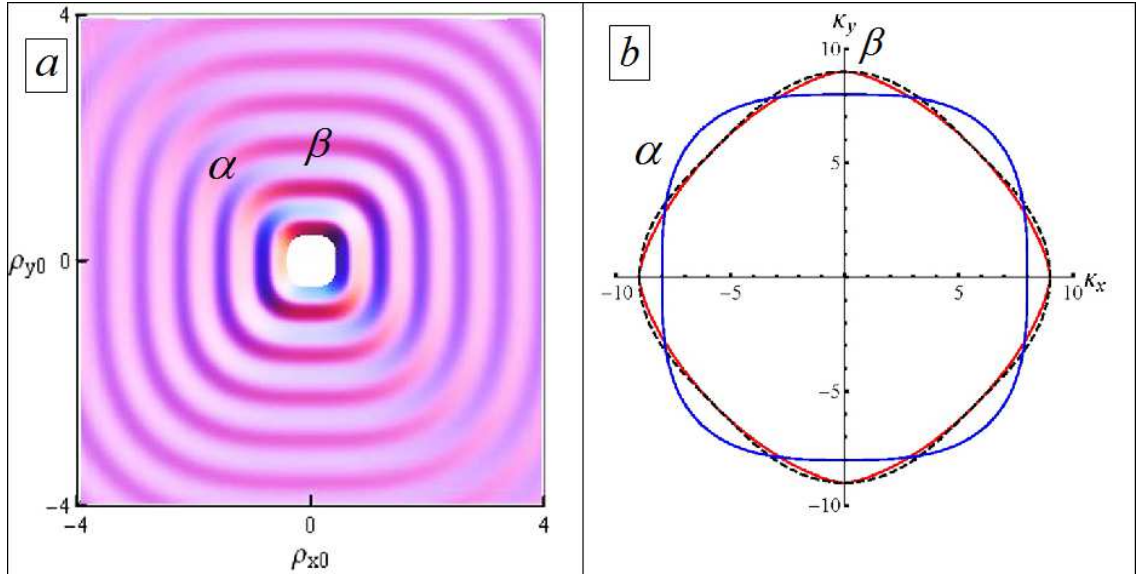}
\caption{$a.$ Interference pattern in the conductance $G$ as obtained from
Eq.~(\protect\ref{G_asympt}) resulting from scattering of the electrons by a
surface defect. The coordinates $\protect\rho _{x0}$ and $\protect\rho _{y0}$
are given in units of $1/k_{0}$. The support function is given by Eq.~(%
\protect\ref{h_kw}). $b.$ Plots of the Fermi contour (solid), its support
function $h$ (short-dashed) and $1/h$ (long-dashed) , with $\protect\kappa %
_{x}$ and $\protect\kappa _{y}$ in units of $k_{0}$. }
\label{fig4}
\end{figure}

Figure~4a illustrates the standing wave pattern in the conductance $G\left(
\mathbf{\rho }_{0}\right) $ (\ref{G_asympt}) around the defect for a model
FC, which we take to be a convex curve described by the support function
\cite{Rabinov}%
\begin{equation}
h\left( \phi \right) =k_{0}\left( \cos ^{2}2\phi +8\right) .  \label{h_kw}
\end{equation}%
In absence of spin-orbit interaction the 2D Fermi surface has a centre of
symmetry $\varepsilon \left( \mathbf{\kappa }\right) =\varepsilon \left( -%
\mathbf{\kappa }\right) $ and also the contour described by the support
function $h\left( \phi \right) $ (\ref{h_kw}) acquires this property, $%
h(\phi )=h\left( \phi +\pi \right) .$ The parametric equations of the curve
in Fig.4 can be easily found from Eqs. (\ref{kx}) and (\ref{ky}). Figure~4b
shows the difference between the true form of the curve and its support
function.

The relation between contours of constant phase $\rho _{0}=\mathrm{const.}%
/2h(\phi )$ in the oscillatory pattern of the conductance (\ref{G_asympt})
in Fig.~5a and the FC in Fig.~5b, is can be understood as follows \cite%
{Avotina06}. For an anisotropic FC the surface electrons move along the
direction of the velocity vector $\mathbf{v}_{\mathbf{\kappa }}$, which need
not generally be parallel to the wave vector $\mathbf{\kappa .}$ The
standing wave at any point of the STM image is defined by the velocity
directed from the contact to the defect. For parts of the FC having a small
curvature (illustrated by the point $\alpha $ in Fig.~5b) all electrons for
different $\mathbf{\kappa }$ in this region have similar velocities. In real
space together they form a narrow electron beam and contribute to only a
small sector of the STM image. Conversely, for electrons belonging to small
parts of the FC having a large curvature (illustrated by the point $\beta $
in Fig.~5), a small change of the angle $\phi $ (and consequently a small
change of $\mathbf{k}\left( \phi \right) $) results in a large change in the
direction of the velocity. Such small parts of the FC define large sectors
in the interference pattern. We emphasize that, despite the resemblance, the
contours of constant phase in $G\left( \mathbf{\rho }_{0}\right) $ are not
just rotated Fermi contours.

\section{Conclusion.}

In summary, we have investigated the conductance of a small-size tunnel
contact for the case of electron tunneling from surface states into bulk
states of the "tip". Electron scattering by a single surface defect is taken
into account. For an arbitrary shape of the Fermi contour of the 2D surface
states an asymptotically exact formula for the STM conductance is obtained,
in the limit of a high tunnel barrier and large distances between the
contact and the defect. The relation between the standing wave pattern in
the conductance and the geometry of 2D Fermi contour is analyzed. We show
that the real space STM image does not show the Fermi surface directly, but
gives the contours of the inverse support function $1/h$ of the 2D Fermi
contour. A rigorous mathematical procedure for the FC reconstruction from
the real space STM images is described.

Today STM imaging has become a new method of fermiology. By using
Fourier-transform scanning tunnelling spectroscopy the FC may be found from
the standing wave pattern of the electrons near the Fermi energy, caused by
defects in the surface. To establish a correspondence between the observed
contours and the actual FC various theoretical approaches have been proposed
(for a review of experimental and theoretical results on this subject see
\cite{Simon}). However in some cases (see, for example, Fig.2 in \cite%
{Sprunger}) such a correspondence is not obvious. We propose another
approach, which was very fruitful in bulk metal physics \cite{LAK} -
experimental results are compared with theoretical formulas obtained for
arbitrary Fermi surfaces - i.e. the inverse problem of the Fermi surface
reconstruction from the experimental data must be solved. The formulas
obtained in this papers for oscillations of the conductance of the tunnel
point contact around point-like surface defects and the procedure of the FC
reconstruction directly from real space STM image is a more rigorous
alternative to the Fourier transform of STM images.

\section{Appendix I: Solution of the Schr\"{o}dinger equation.}

We search for the solution of Eq.~(\ref{Schrod}) corresponding to electron
tunneling from the surface states at $z>0$ into the bulk states in the
half-space $z<0.$ Hereinafter we follow the procedure for finding the wave
function of transmitted electrons $\psi _{\mathrm{tr}}\left( \mathbf{\rho }%
,z\right) $ in the limits $U_{0}\rightarrow \infty ,$ $g\rightarrow 0$ that
was proposed in Refs.~\cite{Kulik,Avotina05}. The wave function of surface
states $\psi _{\mathrm{sur}}\left( \mathbf{\rho },z\right) $ at $z\geqslant
0 $ we search as a sum
\begin{equation}
\psi _{\mathrm{sur}}\left( \mathbf{r}\right) \simeq \varphi _{0\mathrm{sur}%
}\left( \mathbf{r}\right) +\frac{1}{U_{0}}\varphi _{1\mathrm{sur}}\left(
\mathbf{r}\right) ,  \tag{A1.1}  \label{sur}
\end{equation}%
where the second term is a small perturbation of surface state due to finite
probability of tunneling through the contact. In approximation to zeroth
order in $1/U_{0}$ the electrons cannot tunnel through the barrier and the
function $\varphi _{0\mathrm{sur}}\left( \mathbf{r}\right) $ satisfies the
zero boundary condition%
\begin{equation}
\varphi _{0\mathrm{sur}}\left( \mathbf{\rho },z=0\right) =0.  \tag{A1.2}
\label{sur=0}
\end{equation}%
The wave function of transmitted electrons $\psi _{\mathrm{tr}}\left(
\mathbf{\rho },z\right) $ is not zero to first order in $1/U_{0}$%
\begin{equation}
\psi _{\mathrm{tr}}\left( \mathbf{r}\right) \simeq \frac{1}{U_{0}}\varphi _{1%
\mathrm{tr}}\left( \mathbf{r}\right) .  \tag{A1.3}  \label{tr}
\end{equation}%
Substituting the Eqs. (\ref{sur}), (\ref{tr}) in the boundary conditions (%
\ref{equal}), (\ref{jump_wf}) and equating the terms of the same order in $%
1/U_{0}$ we obtain the boundary conditions,
\begin{equation}
\varphi _{1\mathrm{sur}}\left( \mathbf{\rho },+0\right) =\varphi _{1\mathrm{%
tr}}\left( \mathbf{\rho },-0\right) ,  \tag{A1.4}  \label{contin}
\end{equation}%
\begin{equation}
\left. \frac{\partial }{\partial z}\varphi _{0\mathrm{sur}}\left( \mathbf{%
\rho },z\right) \right\vert _{z=0}=\frac{2m_{z}}{\hbar ^{2}}f\left( \mathbf{%
\rho }\right) \varphi _{1\mathrm{tr}}\left( \mathbf{\rho },0\right) .
\tag{A1.5}  \label{jump}
\end{equation}%
The function $\varphi _{0\mathrm{sur}}\left( \mathbf{\rho },z\right) $ will
be found in linear approximation in the constant $g.$ The unperturbed wave
function (in the zeroth approximation in $1/U_{0}$ and $g$) $\varphi _{00%
\mathrm{sur}}\left( \mathbf{\rho },z\right) $ can be easily found%
\begin{equation}
\varphi _{00\mathrm{sur}}\left( \mathbf{\rho },z\right) =\frac{1}{\sqrt{%
L_{x}L_{y}}}e^{i\mathbf{\kappa \rho }}\chi _{0}\left( z\right) .  \tag{A1.6}
\label{psi0}
\end{equation}%
In Eq.~(\ref{psi0}) $L_{x}\simeq L_{y}$ are the lateral sizes of the
interface $\left( L_{x,y}\rightarrow \infty \right), and $ $\chi _{0}\left(
z\right) $ is the solution to the equation
\begin{equation}
\frac{\hslash ^{2}}{2m_{z}}\frac{\partial ^{2}\chi _{0}\left( z\right) }{%
\partial z^{2}}+\left( \varepsilon _{0}-V_{\mathrm{sur}}\left( z\right)
\right) \chi _{0}\left( z\right) =0,\quad z\geqslant 0,  \tag{A1.7}
\label{eq_hi}
\end{equation}%
subject to the boundary conditions and normalization condition
\begin{gather}
\chi _{0}\left( 0\right) =0,\quad \chi _{0}\left( z\rightarrow \infty
\right) \rightarrow 0,  \tag{A1.8}  \label{bound_cond_hi} \\
\int\limits_{0}^{\infty }dz\left\vert \chi _{0}\left( z\right) \right\vert
^{2}=1,  \notag
\end{gather}%
respectively. We will assume that at $\varepsilon \leqslant \varepsilon _{%
\mathrm{F}}$ only one discrete quantum state $\varepsilon _{0}$ is filled in
the surface potential well (see Fig.2). The solution $\varphi _{00\mathrm{sur%
}}\left( \mathbf{r}\right) $ of (\ref{psi0}) describes the wave function of
the surface states near an ideal impermeable interface. \ The correction to
the wave function (\ref{psi0}) linear in the constant $g$ can be expressed
by means of the Green's function $G^{+}\left( \mathbf{r,r}^{\prime
};\varepsilon \right) $ of the unperturbed surface states \cite{Avotina06}
in the field of the potential $V_{\mathrm{sur}}\left( z\right) $ near the
impenetrable interface.

To leading order in the constant $g$ the functions $\psi _{1}\left( \mathbf{r%
}\right) $ and $\varphi _{1}\left( \mathbf{r}\right) $ can be written as%
\begin{equation}
\varphi _{0\mathrm{sur}}\left( \mathbf{r}\right) =\varphi _{00\mathrm{sur}%
}\left( \mathbf{r}\right) +\varphi _{00\mathrm{sur}}\left( \mathbf{r}%
_{0}\right) g\int d\mathbf{r}^{\prime }D\left( \mathbf{r}^{\prime }-\mathbf{r%
}_{0}\right) G^{+}\left( \mathbf{r,r}^{\prime };\varepsilon \right) .
\tag{A1.9}  \label{psi1phi1}
\end{equation}

The retarded Green's function of the surface states is given by
\begin{equation}
G^{+}\left( \mathbf{r,r}^{\prime };\varepsilon \right) =\chi _{0}\left(
z\right) \chi _{0}^{\ast }\left( z^{\prime }\right) G_{2D}^{+}\left( \mathbf{%
\rho -\rho }^{\prime };\varepsilon -\varepsilon _{0}\right)  \tag{A1.10}
\label{GreenF}
\end{equation}%
with
\begin{equation}
G_{2D}^{+}\left( \mathbf{\rho };\varepsilon \right) =\frac{1}{\left( 2\pi
\right) ^{2}}\int\limits_{-\infty }^{\infty }d^{2}q\frac{e^{i\mathbf{q\rho }}%
}{\varepsilon -\varepsilon _{2D}\left( \mathbf{q}\right) +i0}.  \tag{A1.11}
\label{G2D}
\end{equation}

The wave function for the electrons that are transmitted through the barrier
$\psi _{\mathrm{tr}}\left( \mathbf{r}\right) $ can be found along the lines
described in Refs.~\cite{Kulik,Avotina05}. Taking the Fourier transform of
the unknown function $\psi _{\mathrm{tr}}\left( \mathbf{r}\right) $ in the
half-space $z<0$ (Fig.~\ref{fig1}),
\begin{equation}
\psi _{\mathrm{tr}}\left( \mathbf{\rho },z\right) =\int\limits_{-\infty
}^{\infty }d\mathbf{\kappa }^{\prime }e^{i\mathbf{\kappa }^{\prime }\mathbf{%
\rho }}\Phi \left( \mathbf{\kappa }^{\prime },z\right) ,  \tag{A1.12}
\label{phiFour}
\end{equation}%
and substituting this in the Schr\"{o}dinger equation,
\begin{equation}
\left( \varepsilon _{2D}\left( \frac{\hbar \partial }{i\partial \mathbf{\rho
}}\right) +\frac{\hslash ^{2}\partial ^{2}}{2m_{z}\partial z^{2}}\right)
\psi _{\mathrm{tr}}\left( \mathbf{r}\right) +\varepsilon \psi _{\mathrm{tr}%
}\left( \mathbf{r}\right) =0,\quad z<0,  \tag{A1.13}  \label{Schrod_z<0}
\end{equation}%
we find for the Fourier component $\Phi \left( \mathbf{\kappa }^{\prime
},z\right) $ a solution corresponding to a propagating wave along the $z$
direction,
\begin{equation}
\Phi \left( \mathbf{\kappa }^{\prime },z\right) =\Phi \left( \mathbf{\kappa }%
^{\prime },0\right) \exp \left( -ik_{z}^{\prime }z\right) ,\qquad \qquad
\qquad z\leqslant 0,  \tag{A1.14}  \label{Phi(k,z)}
\end{equation}%
with $k_{z}^{\prime }=\sqrt{2m_{z}\left( \varepsilon -\varepsilon
_{2D}\left( \mathbf{\kappa }^{\prime }\right) \right) }/\hbar.$ In order to
obtain the waves diverging from the contact and to satisfy the boundary
condition (\ref{-inf}) we must take $\func{Im}k_{z}^{\prime }<0$ at $%
\varepsilon _{2D}\left( \mathbf{\kappa }^{\prime }\right) >\varepsilon $.
From the simplified boundary condition (\ref{jump}), with known wave
function $\varphi _{0\mathrm{sur}}\left( \mathbf{r}\right) $ (\ref{psi0}) to
zeroth approximation in the constant $g$, one can find the function $\varphi
_{1\mathrm{tr}}\left( \mathbf{\rho },0\right) $ in the plane of interface $%
z=0.$ Relation (\ref{psi1phi1}) gives us $\varphi _{1\mathrm{tr}}\left(
\mathbf{\rho },0\right) $ to first approximation in the small constant $g$,
\begin{gather}
\varphi _{1\mathrm{tr}}\left( \mathbf{\rho },0\right) =-\frac{\hbar ^{2}}{%
2m_{z}f\left( \mathbf{\rho }\right) \sqrt{L_{x}L_{y}}}e^{i\mathbf{\kappa
\rho }}\chi _{0z}^{\prime }\left( +0\right) \left[ 1+\right.  \tag{A1.15}
\label{phi1tr} \\
\left. g\int\limits_{-\infty }^{\infty }d\mathbf{\rho }^{\prime
}\int\limits_{0}^{\infty }dz^{\prime }D_{0}\left( \mathbf{\rho }^{\prime }%
\mathbf{-\rho }_{0},z^{\prime }\right) \left\vert \chi _{0}\left( z^{\prime
}\right) \right\vert ^{2}e^{i\mathbf{\kappa \rho }^{\prime
}}G_{2D}^{+}\left( \mathbf{\rho -\rho }^{\prime };\varepsilon -\varepsilon
_{0}\right) \right]  \notag
\end{gather}%
The inverse Fourier transform allows us to express $\ \Phi \left( \mathbf{%
\kappa }^{\prime },0\right) $ in terms of the known function $\varphi \left(
\mathbf{\rho },0\right) $
\begin{equation}
\Phi \left( \mathbf{\kappa }^{\prime },0\right) =\frac{1}{(2\pi )^{2}}%
\int\limits_{-\infty }^{\infty }d\mathbf{\rho }^{\prime }e^{-i\mathbf{\kappa
}^{\prime }\mathbf{\rho }^{\prime }}\varphi \left( \mathbf{\rho }^{\prime
},0\right) ,  \tag{A1.16}  \label{Phi(k,0)}
\end{equation}%
and we finally obtain the wave function for the transmitted electrons
\begin{equation}
\psi _{\mathrm{tr}}\left( \mathbf{\rho },z\right) =\frac{1}{(2\pi )^{2}U_{0}}%
\int\limits_{-\infty }^{\infty }d\mathbf{\rho }^{\prime
}\int\limits_{-\infty }^{\infty }d\mathbf{\kappa }^{\prime }\varphi _{1%
\mathrm{tr}}\left( \mathbf{\rho }^{\prime },0\right) e^{i\mathbf{\kappa }%
^{\prime }\left( \mathbf{\rho -\rho }^{\prime }\right) -ik_{z}^{\prime
}\left\vert z\right\vert },  \tag{A1.17}  \label{psi_tr}
\end{equation}%
were $\varphi _{1\mathrm{tr}}\left( \mathbf{\rho }^{\prime },0\right) $ is
given by Eq. (\ref{phi1tr}).

\section{Appendix II: Asymptotes for $\protect\rho \rightarrow \infty $ of
the Green function $G_{2D}^{+}\left( \mathbf{\protect\rho };\protect%
\varepsilon \right) $ of 2D electrons with an arbitrary Fermi contour.}

After replacing the integration over the 2D vector $\mathbf{q}$ by
integrations over the energy $\varepsilon ^{\prime }=$ $\varepsilon
_{2D}\left( \mathbf{q}\right) $ and over the arc length $l_{\mathbf{q}}$ of
the constant energy contour, Eq.~(\ref{G2D}) takes the form%
\begin{equation}
G_{2D}^{+}\left( \mathbf{\rho };\varepsilon \right) =\frac{1}{\left( 2\pi
\right) ^{2}}\int\limits_{0}^{\infty }\frac{\Lambda \left( \varepsilon
^{\prime },\mathbf{\rho }\right) d\varepsilon ^{\prime }}{\varepsilon
-\varepsilon ^{\prime }+i0},  \tag{A2.1}  \label{G2D_1}
\end{equation}%
where
\begin{equation}
\Lambda \left( \varepsilon ^{\prime },\mathbf{\rho }\right)
=\oint\limits_{\varepsilon ^{\prime }=\varepsilon _{2D}\left( \mathbf{q}%
\right) }\frac{dl_{\mathbf{q}}}{\hbar v_{\mathbf{q}}}e^{i\mathbf{q\rho }},
\tag{A2.2}  \label{L1}
\end{equation}%
and $v_{\mathbf{q}}=\left\vert \partial \varepsilon _{2D}\left( \mathbf{q}%
\right) /\hbar \partial \mathbf{q}\right\vert $ is the absolute value of the
2D velocity vector. For $\rho \rightarrow \infty $ the integral in Eq. (\ref%
{L1}) can by calculated asymptotically by using the stationary phase method
(see, for example, Ref.~\cite{Fedoriuk}). Let us parameterize the curve $%
\varepsilon _{2D}\left( \mathbf{q}\right) =\varepsilon ^{\prime }$ by using
the angle $\phi $ in the $q_{x}q_{y}$-plane as a parameter, $%
q_{x,y}=q_{x,y}\left( \varepsilon ^{\prime },\phi \right).$ The element of
arc length $dl_{\mathbf{q}}$ can then be expressed as $dl_{\mathbf{q}}=\sqrt{%
\dot{q}_{x}^{2}+\dot{q}_{y}^{2}}d\phi $ and we obtain%
\begin{equation}
\Lambda _{as}\left( \varepsilon ^{\prime },\mathbf{\rho }\right) \simeq
\frac{1}{\hbar v_{\mathbf{q}}}\sqrt{\frac{2\pi \left( \dot{q}_{x}^{2}+\dot{q}%
_{y}^{2}\right) }{\left\vert \ddot{q}_{x}\rho _{x}+\ddot{q}_{y}\rho
_{y}\right\vert }}\left. \exp \left[ i\mathbf{q\rho }+\frac{\pi i}{4}\text{%
sgn}\left( \ddot{q}_{x}\rho _{x}+\ddot{q}_{y}\rho _{y}\right) \right]
\right\vert _{\phi =\phi _{st}}+O\left( \frac{1}{\rho }\right) ,  \tag{A2.3}
\label{G2D_asym}
\end{equation}%
where the dot over a function denotes differentiation with respect to $\phi
. $ The stationary phase point $\phi =\phi _{st}\left( \varepsilon ^{\prime
}\right) $ is defined by the equation%
\begin{equation}
\left. \dot{q}_{x}\rho _{x}+\dot{q}_{y}\rho _{y}\right\vert _{\phi =\phi
_{st}}=0.  \tag{A2.4}  \label{stphpoint_1}
\end{equation}%
Note that the total derivative of the energy $\varepsilon _{2D}\left(
\mathbf{q}\right) =\varepsilon $ with respect to $\phi $ is equal to zero
because this energy is the same for all directions $\phi $ of the vector $%
\mathbf{q}$,
\begin{equation}
\dot{\varepsilon}_{2D}\left( \mathbf{q}\right) =\hbar v_{x}\dot{q}_{x}+\hbar
v_{y}\dot{q}_{y}=0.  \tag{A2.5}  \label{eps_point}
\end{equation}%
Equation~(\ref{eps_point}) provides a relation between the derivatives $\dot{%
q}_{x,y}$ and the components of the velocity $v_{x,y}.$ Introducing the
curvature $K\left( \mathbf{q}\right) $ of the constant energy contour $%
\varepsilon _{2D}\left( \mathbf{q}\right) =\varepsilon ^{\prime }$
\begin{equation}
K\left( \mathbf{q}\right) =\frac{\ddot{q}_{y}\dot{q}_{x}-\ddot{q}_{x}\dot{q}%
_{y}}{\left( \dot{q}_{x}^{2}+\dot{q}_{y}^{2}\right) ^{3/2}},  \tag{A2.6}
\label{K(q)}
\end{equation}%
Eqs.~(\ref{G2D_asym}) and (\ref{stphpoint_1}) can be rewritten in the form
\begin{equation}
\Lambda _{as}\left( \varepsilon ^{\prime },\mathbf{\rho }\right) \simeq
\sqrt{2\pi }\left. \frac{\exp \left[ i\mathbf{q\rho }+\frac{\pi i}{4}\text{%
sgn}K\left( \mathbf{q}\right) \right] }{\hbar v_{\mathbf{q}}\sqrt{\rho
\left\vert K\left( \mathbf{q}\right) \right\vert }}\right\vert _{\mathbf{q}=%
\mathbf{q}_{st}}+O\left( \frac{1}{\rho }\right) ,  \tag{A2.7}
\label{GreenFasym}
\end{equation}%
and%
\begin{equation}
\left. \frac{v_{x}}{v_{y}}\right\vert _{\mathbf{q}=\mathbf{q}_{st}}=\frac{%
\rho _{x}}{\rho _{y}},  \tag{A2.8}  \label{stphpoint_2}
\end{equation}%
where $\mathbf{q}_{st}=\left( q_{x}\left( \phi _{st}\right) ,q_{y}\left(
\phi _{st}\right) \right) $. The equality (\ref{stphpoint_2}) is satisfied
when the velocity $\mathbf{v}_{\mathbf{q}_{st}}$ is parallel or antiparallel
to the vector $\mathbf{\rho }.$ We choose the solution of Eq. (\ref%
{stphpoint_2}) with $\mathbf{v}_{\mathbf{q}_{st}}\parallel $ $\mathbf{\rho }$
that corresponds to the outgoing waves. Generally, for arbitrarily
complicated (non-convex) constant energy contours there can be many
solutions $\mathbf{q}_{st}^{\left( s\right) }$ $\left( s=1,2,...\right) $
and in Eq.~(\ref{GreenFasym}) one must sum over all of them.

\begin{figure}[tbp]
\includegraphics[width=10cm,angle=0]{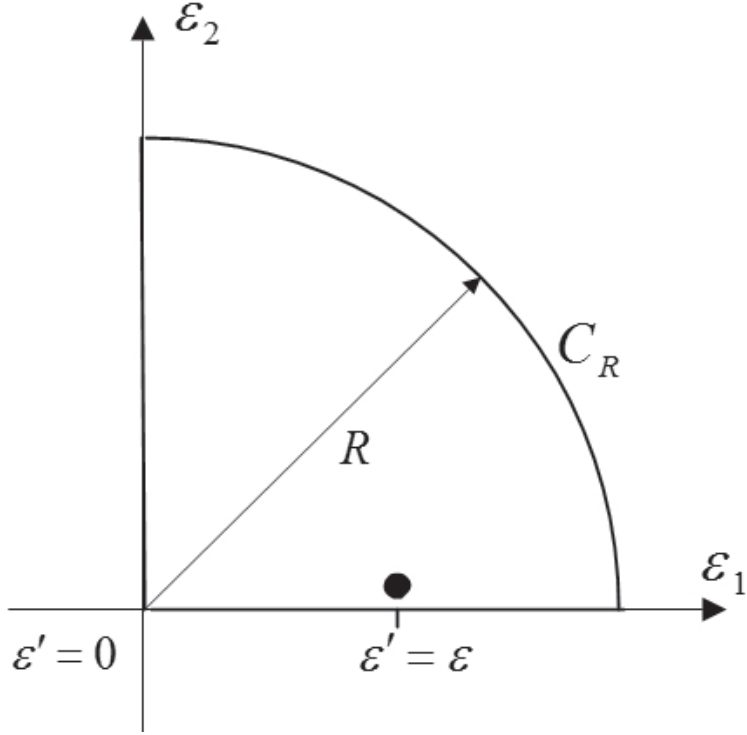}
\caption{Contour of integration used in Eqs.~(\protect\ref{J1}) and (\protect
\ref{J3}). The black dot shows the position of the pole of the integrand. }
\label{figx}
\end{figure}
In order to calculate the integral over $\varepsilon ^{\prime }$ in Eq.~(\ref%
{G2D_1}) we consider the integral $J_{C}$ along the closed contour $C$ shown
in Fig.~\ref{figx},%
\begin{equation}
J_{C}=\frac{1}{\left( 2\pi \right) ^{3/2}}\lim_{R\rightarrow \infty
}\int\limits_{C}\frac{\Lambda _{as}\left( \varepsilon ^{\prime },\mathbf{%
\rho }\right) d\varepsilon ^{\prime }}{\varepsilon -\varepsilon ^{\prime }+i0%
}.  \tag{A2.9}  \label{J1}
\end{equation}%
There is only one pole $\varepsilon =\varepsilon ^{\prime }+i0$ inside $C$
and this integral is equal to
\begin{equation}
J_{C}=\lim_{R\rightarrow \infty }\left\{
\int\limits_{0}^{R}+\int\limits_{C_{R}}+\int\limits_{iR}^{0}d\varepsilon
^{\prime }\right\} =2\pi i\Lambda _{as}\left( \varepsilon ,\mathbf{\rho }%
\right) .  \tag{A2.10}  \label{J2}
\end{equation}%
The first integral \ in (\ref{J2}) for $R\rightarrow \infty $ is the desired
integral in Eq.~(\ref{G2D_1}). The second integral along the arc $C_{R}$
vanishes for $R\rightarrow \infty $ if Re$\left( i\mathbf{q\rho }\right) <0$
in the first quadrant of the plane of the complex variable $\varepsilon
^{\prime }=\varepsilon _{1}+i\varepsilon _{2}$. The third integral in (\ref%
{J2}) along the complex axis $i\varepsilon _{2}$ rapidly decreases with
increasing distance $\rho $, more rapidly than the first one because of the
exponential dependence of the integrand.

The last two statements can be proven explicitly for an isotropic dispersion
law of 2D charge carriers. For a circular contour of constant energy $%
\varepsilon =\left( \hbar \kappa \right) ^{2}/2m,$ $\varepsilon ^{\prime
}=\left( \hbar q^{\prime }\right) ^{2}/2m,$ $v=\hbar q^{\prime }/m,$ $%
K\left( \mathbf{q}\right) =1/q^{\prime },$ $\mathbf{q}_{st}\mathbf{\rho }%
=q^{\prime }\rho $. Replacing the integration over $\varepsilon ^{\prime }$
by \ integration over $q^{\prime }$ we obtain
\begin{equation}
J_{C}=\lim_{R\rightarrow \infty }\left\{
\int\limits_{0}^{R}+\int\limits_{C_{R}}+\int\limits_{iR}^{0}dq^{\prime
}\right\} =\frac{m}{\pi \hbar ^{2}\sqrt{\rho }}\int\limits_{C}\frac{\sqrt{%
q^{\prime }}dq^{\prime }}{\kappa ^{2}-q^{\prime 2}+i0}\exp \left[ iq^{\prime
}\rho +\frac{\pi i}{4}\right] .  \tag{A2.11}  \label{J3}
\end{equation}%
For the integral (\ref{J3}) we use the same contour as for integral (\ref{J1}%
) (see Fig.~\ref{figx}). Let us replace the integration variable $q^{\prime
} $ in the second integral along the circle quarter $C_{R}$ by $q^{\prime
}=R\cdot e^{i\chi }.$ Then it is easy to estimate the absolute value of the
integral as,
\begin{gather}
\left\vert \int\limits_{C_{R}}\frac{\sqrt{q^{\prime }}dq^{\prime }}{\kappa
^{2}-q^{\prime 2}+i0}\exp \left[ iq^{\prime }\rho +\frac{\pi i}{4}\right]
\right\vert <\frac{1}{\sqrt{R}}\int\limits_{0}^{\pi /2}d\chi e^{-R\rho \sin
\chi }<  \tag{A2.12} \\
\frac{1}{\sqrt{R}}\int\limits_{0}^{\pi /2}d\chi e^{-2R\rho \chi /\pi }=\frac{%
\pi }{2\rho R^{3/2}}\left( 1-e^{-R\rho }\right) \underset{R\rightarrow
\infty }{\rightarrow }0.  \notag
\end{gather}%
After substituting $\xi =-iq^{\prime }$ the third integral along imaginary
axis takes the form
\begin{gather}
\lim_{R\rightarrow \infty }\frac{1}{\sqrt{\rho }}\int\limits_{iR}^{0}\frac{%
\sqrt{q^{\prime }}dq^{\prime }}{\kappa ^{2}-q^{\prime 2}+i0}\exp \left[
iq^{\prime }\rho +\frac{\pi i}{4}\right] =\frac{1}{\sqrt{\rho }}%
\int\limits_{0}^{\infty }\frac{\sqrt{\xi }e^{-\xi \rho }d\xi }{\kappa
^{2}+\xi ^{2}}=  \label{A2.13} \\
\frac{\pi }{\sqrt{\kappa \rho }}\left[ \cos \kappa \rho \left( 1-2C\left(
\sqrt{\kappa \rho }\right) \right) +\sin \kappa \rho \left( 1-2S\left( \sqrt{%
\kappa \rho }\right) \right) \right] \underset{\rho \rightarrow \infty }{%
\longrightarrow }\frac{\sqrt{\pi }}{2\kappa ^{2}\rho ^{2}}+O\left( \frac{1}{%
\rho ^{3}}\right) ,  \notag
\end{gather}%
where $C\left( z\right) $ and $S\left( z\right) $ are the Fresnel integrals%
\begin{equation}
\left\{
\begin{array}{c}
C\left( z\right) \\
S\left( z\right)%
\end{array}%
\right\} =\sqrt{\frac{2}{\pi }}\int\limits_{0}^{z}dt\left\{
\begin{array}{c}
\cos t^{2} \\
\sin t^{2}%
\end{array}%
\right\} .  \tag{A1.14}
\end{equation}%
Finally we obtain the following asymptotic expression for the Green function%
\begin{equation}
G_{2D}^{+}\left( \mathbf{\rho };\varepsilon \right) \simeq \frac{i}{\sqrt{%
2\pi }}\left. \frac{\exp \left[ i\mathbf{q\rho }+\frac{\pi i}{4}\text{sgn}%
K\left( \mathbf{q}\right) \right] }{\hbar v_{\mathbf{q}}\sqrt{\rho
\left\vert K\left( \mathbf{q}\right) \right\vert }}\right\vert _{\mathbf{q}=%
\mathbf{q}_{st}\left( \varepsilon ,\phi _{st}\right) },\quad \rho
\rightarrow \infty .  \tag{A1.15}  \label{G2D_as}
\end{equation}

\end{document}